\documentclass[aps,prl,preprint,groupedaddress]{revtex4-2}

\usepackage{amsmath}
\usepackage{amssymb}
\usepackage[pdftex]{graphicx}

\usepackage{tabularx}
\usepackage{footmisc}
\usepackage{hyperref}

\begin{document}


\title{In Situ Dynamic Four-Dimensional Strong Field Ionization Tomography}


\author{Noam Shlomo, Eugene Frumker}
\email[]{efrumker@bgu.ac.il}
\affiliation{Department of Physics, Ben-Gurion University of the Negev, Beer-Sheva 84105, Israel}


\begin{abstract}

We  present an approach for in situ dynamic four-dimensional (4D=3D space + 1D time) laser induced strong field ionization tomography, particularly suited for measuring far-from-equilibrium systems, such as supersonic and hypersonic pulsed gas jets.
The inherent physical nature of strong-field interaction leads to significantly enhanced spatial resolution, with unique intensity-resolution coupling and localization of measurement depth. 
By employing femtosecond laser, temporal resolution is primarily limited by electronic jitter, typically to a few tens of picoseconds.

\end{abstract}

\maketitle












The space-time dynamics of supersonic and hypersonic jets are of fundamental and practical significance across various scientific and engineering disciplines.
These include, for example,  laser particle acceleration (LPA) \cite{Modena_Electron_nature_1995}, attosecond science \cite{Itatani_tomog_Nature2004, Kienberger_Atomic_transient_recorder_2004, Li_dynamics_N2O4_Science_2008}, inertial confinement fusion (ICF) \cite{Denavit__gas_targets_PhysPlasmas_1994}, x-ray sources \cite{Failor_fluorescence_gas_tomogr_RSI_2003},  and cold chemistry \cite{Thorpe_tomography_CPL_2009,  Henson_Observation_resonances_beams_Science_2012}.

Systems exhibiting highly non-stationary supersonic dynamics, such as pulsed jets, are particularly intriguing and important.  
These jets play a crucial role in attosecond science, as a target medium for high harmonic generation (HHG) in atomic gases \cite{Bellini_temp_coherence_PRL1998}, as well as facilitating experiments involving aligned \cite{Itatani_tomog_Nature2004} and oriented \cite{Frumker_orientation_HHG_PRL2012,frumker_wavepacket_PRL2012} molecules. Moreover, they find extensive use in atomic and molecular beam physical chemistry \cite{even_cooling1K_JCP_2000, Narevicius_Stopping_beams_PRL_2008} and other related fields.

Theoretical modeling of such far-from-equilibrium systems remains a significant challenge and an active research area. For many cases of interest, adequate theoretical models are still unavailable.
Therefore, an experimental tool enabling direct space-time observation and characterization of supersonic jets with high spatial and temporal resolution is highly desirable and critical from both fundamental research and practical application perspectives.
 
Several experimental efforts to characterize spatial and temporal properties of the gas jets have been reported. These include static 2D and 3D spatial tomographic reconstruction techniques based on linear or perturbative non-linear interactions \cite{malka2000charact, Failor_fluorescence_gas_tomogr_RSI_2003, Thorpe_tomography_CPL_2009, Sschofield_Absolute_density_tomography_RSI_2009, Wachulak_Extreme_UV_Tomog_AppPhysB_2014, Tejeda_mapping_supersonic_PRL_1996, Comby_Absolute_Gas_Density_OE_2018}, along with 3D laser induced strong-field ionization gas jet tomography \cite{Tchulov_LaserStrongFieldTom_ScRep_2017}.  Studies employing  the HHG absorbtion spectroscopy \cite{Hagmeister_HHG_Characterization_Jet_AppPhysB_2022} and Nomarski interference  \cite{Liu_SpaceTime_Jet_FiP_2023} demonstrated  a one dimensional temporal measurement along the jet axis.
Interferometric phase measurements \cite{Nagyilles_time_resolved_PysRevAppl_2023} of the gas jet have recently been reported; however, the dynamic range in the interferometric measurements is inherently limited and exhibits the ambiguity associated with a modulus $2\pi$ phase shift.

In this paper we introduce a new scheme for comprehensive  space-time (3D+1D) density measurement of jet expansion.
We use this approach to demonstrate dynamic four-dimensional measurement and tomographic reconstruction of supersonic expansion, utlizing pulsed gas jet.
To the best of our knowledge, this represents the most comprehensive experimental measurement of space-time supersonic jet dynamics to date.

The spatial (3D) aspect of our approach leverages the laser-induced strong-field gas jet ionization tomography \cite{Tchulov_LaserStrongFieldTom_ScRep_2017}, thereby inheriting its distinctive features and advantages. 
The fundamental properties of this approach are systematically discussed in  a theoretical paper \cite{Frumker_thory_SFI_tomography_2024}.

Compared to linear tomographic techniques, the nonlinearity of strong-field ionization leads to an order-of-magnitude improvement in volumetric resolution for typical high harmonic generation (HHG) experiment intensities \cite{Frumker_thory_SFI_tomography_2024}. The physical characteristics of this nonlinear interaction are responsible for another advantage: While the depth of focus in linear tomography is constrained by resolution degradation stemming from the divergence of the driving laser beam, in our approach, the depth of scanning is localized \cite{Frumker_thory_SFI_tomography_2024} as the strong-field ionization rate drops rapidly with the divergence of the driving laser beam.
This localization phenomenon is especially useful in eliminating the background signals from a buffer gas or any undesirable gaseous environment.

 In  imaging modalities based on perturbative nonlinear interactions with focused Gaussian laser beams, such as second harmonic generation (SHG) \cite{Gannaway_Scanning_Micr_SHG_OQE_1978}, two-photon fluorescence (TPF) \cite{Denk_TPF_microscopy_Science_1990}, third harmonic generation (THG) \cite{Barad_THG_microscopy_APL1997} or stimulated Raman Scattering (SRS) \cite{Freudiger_SRS_Microscopy_Science_2008}, the enhancement in resolution is intensity-independent and solely depends on the order, $n$, of the nonlinear process, scaling as $\sim1/\sqrt{n}$.  In contrast, in the realm  of strong field nonlinear ionization, there is an inherent physical mechanism that couples the intensity and the resolution in a continuous fashion \cite{Tchulov_LaserStrongFieldTom_ScRep_2017, Frumker_thory_SFI_tomography_2024}.

\begin{figure}
    \begin{center}
	
\includegraphics[width=\linewidth]{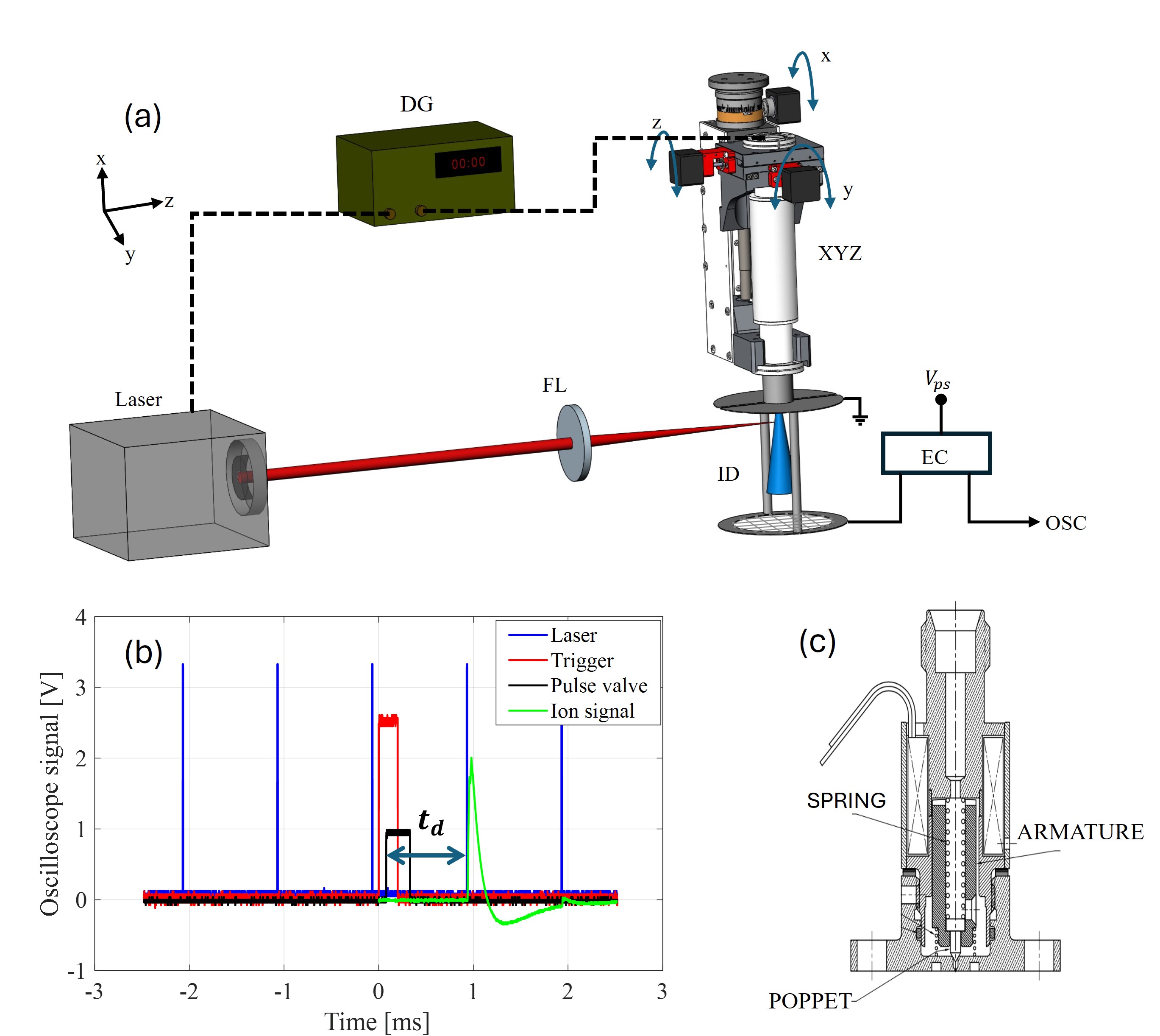}


    \caption{ \label{SetUp_Lebel} (a) Schematics of the experiment. A femtosecond laser beam is focused by a spherical f = 500 mm focusing mirror (FL, shown as a lens for clearity) onto a gas jet. The motorized vacuum manipulator (XYZ) allows for scanning of the gas jet relative to the laser beam.  
      (b) Time sequence: Blue comb depicts the 1KHZ amplified laser pulses sequence, control trigger TTL pulse generated by the delay pulse generator shown in red; actual timing (starting at $t_d$ relative to ionizing laser pulse), of the opening pulsed valve signal shown in black - opening time set to 250 $\mu s$, and the signal delayed by 80 $\mu s$ relatively to the control trigger pulse; green curve shows measured ion signal trace  (c) Schematic of the Parker’s general pulsed valve used in the experiment.}

    \end{center}
\end{figure}

In the nutshell, the amplified femtosecond laser beam  is focused into a supersonic gas jet as shown in Fig. \ref{SetUp_Lebel} (a).
This laser beam is raster scanned across the XY plane to gather tomographic signal data for reconstructing the density map of the gas jet. For the proof-of-principle experiment, we used a Parker’s general pulsed valve, schematically shown in Fig. \ref{SetUp_Lebel} (c), commonly employed in attosecond science \cite{Itatani_tomog_Nature2004, Vozzi_gen_tomography_NatPhys_2011, Frumker_orientation_HHG_PRL2012}, strong field science \cite{Litvinyuk_N2_ion_PRL_2003}, as well as other areas of science and engineering \cite{Shachar_parker_valve_JPhysB_2021}. Equipped with a straight  100 $\mu m$ nozzle, this pulsed valve possess rotational symmetry, so that the XY scan is sufficient to collect a complete data set for the tomographic reconstruction. In the general case of an asymmetric jet, data collection would necessitate rotational scanning of the jet relative to the driving laser beam.

To add the fourth, temporal dimension, we establish control of the ultrafast ionizing laser pulse arrival relative to the gas injection timing.
This is achieved by using digital delay/pulse generator (DG) that allows control and triggering of the injection timing of the pulsed gas jet into the interaction chamber, as depicted in Fig. \ref{SetUp_Lebel}(a).
Scanning both the injection timing, $t_d$, relative to the ionizing laser pulse across the time window of interest, according to the time sequence illustrated in Fig. \ref{SetUp_Lebel}(b),  and the laser beam position relative to the pulsed valve allows us to capture a complete picture of the gas expansion process in space and time.

The temporal resolution of our approach is limited by the larger of the laser pulse duration and the pulsed delay jitter of the delay generator. Our laser pulse duration, on the order of a few tens of femtoseconds, can be considered a delta-function in time on the timescale of supersonic gas expansion dynamics. 
 it is significantly shorter than the electronic jitter of the delay generator, which is around 25 ps and defines the actual temporal resolution.

We measure the temporal profile of the ion signal \cite{Shlomo_In_situ_arXiv_2024} with an oscilloscope (OSC) and use a professional analog electronic circuit simulator, LTspice, to analyze the electronic circuit (EC) and deduce the current flow through the ion detector (ID) as a function of time. By integrating the current, we calculate the total charge generated within the interaction region. With knowledge of the measured beam profile, gas density distribution, and total charge generated, we derive the laser-generated ion/free electron density and absolute gas density in the interaction region \cite{Shlomo_In_situ_arXiv_2024}.

Currently, the dynamic range of our measurement spans over four orders of magnitude, limited by the electronic circuitry (EC) (details of the deployed electronics can be found in \cite{Shlomo_In_situ_arXiv_2024}).
Measurement sensitivity, currently constrained by passive electronics, can be further enhancement through the integration of more sophisticated readout electronics, such as transimpedance amplifiers \cite{Sackinger_transimpedance_amplifiers_book_2017}.

It's noteworthy that the spatial resolution, as defined by ionization width (31 $\mu m$), was determined  by the laser beam width of 37 $\mu m$  and the laser intensity of $8.6\times10^{14}$ $\textrm{W}/\textrm{cm}^2$. 
Note that decreasing the laser intensity enhances resolution at the expense of signal reduction \cite{Frumker_thory_SFI_tomography_2024}. For example, reducing the laser intensity by half will result in
ionization width of 23 $\mu m$, but the signal will drop by 84\%. Therefore, in each specific measurement scenario, the laser intensity can be optimized to strike the optimal balance between resolution and signal-to-noise ratio (SNR).

Our approach offers a distinct advantage over temporal gating methods that rely on short exposure times of the imaging device  \cite{Nagyilles_time_resolved_PysRevAppl_2023}.
Unlike temporal gating, where higher time resolution necessitates shorter exposure times, leading to a weaker signal, our approach achieves high time resolution without compromising the signal level. Consequently, our method allows for significantly improved temporal resolution while maintaining a favorable SNR in measurements.




 Under normal conditions, the spring-loaded poppet (Fig. \ref{SetUp_Lebel}(c))  remains closed, preventing gas flow. A voltage pulse, typically lasting hundreds of microseconds, is applied to the coil assembly to initiate gas injection. This pulse actuates the armature connected to the poppet, overcoming the main spring force and enabling gas flow.
 
The resulting space-time dynamics of the supersonic gas jet is determined by a complex interplay of factors, including the nozzle and poppet shapes, their relative movement, backpressure, gas temperature, and gas type.
Given the rotational symmetry of the jet, scanning the jet in the (XY) plane and the injection timing  is sufficient to acquire a tomographic dataset for complete spatiotemporal reconstruction.

\begin{figure}
    \begin{center}
	
\includegraphics[width=\linewidth]{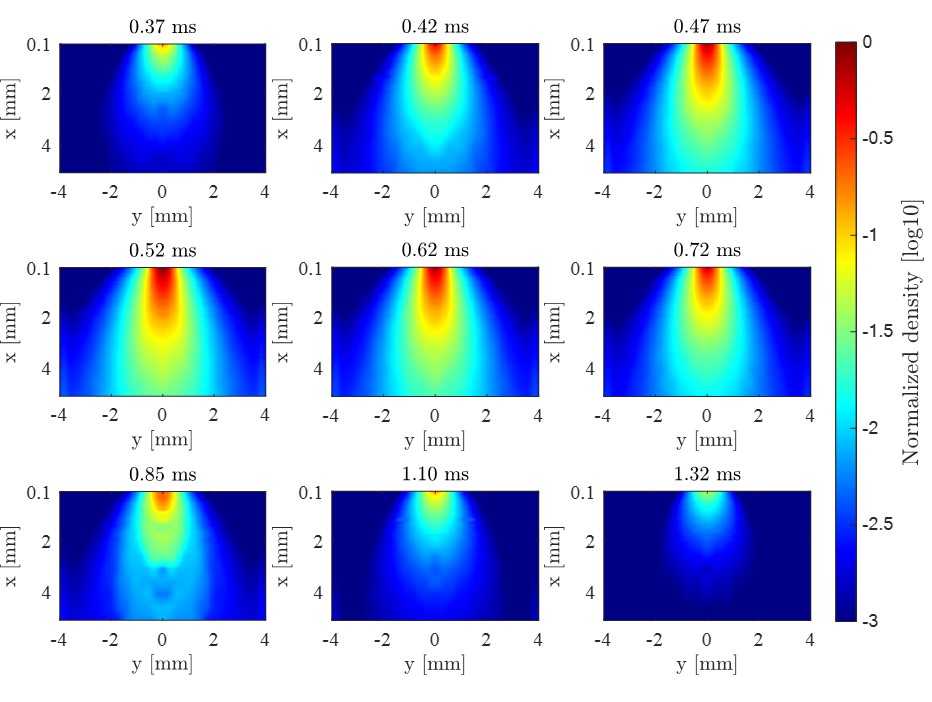}

    \caption{ \label{Fig_Time_Evol_XZtomography}  Supersonic argon gas jet expansion measured using the in situ dynamic  4D laser-induced strong field ionization tomography. Experimental conditions included a gas backpressure of 7.5 bar and a femtosecond laser pulse energy of 1 mJ. The pressure in the vacuum chamber was maintained at approximately $10^{-7}$ Torr prior to injection. Slices along the gas jet propagation are shown at various time delays relative to the start of injection. The density is normalized to the maximum measured value of $9\times 10^{15} \mathrm{cm}^{-3}$. See Supplemental Material at [URL will be inserted by publisher \url{https://drive.google.com/file/d/1igiLQWtlnmJN8s_bu_dKz5KkUZot1akU/view?usp=drive_link} ] for the space-time movie.}
    \end{center}
\end{figure}

Figure \ref{Fig_Time_Evol_XZtomography} displays snapshots of the (XY) cross-section of the measured argon gas jet's spatiotemporal dynamics after tomographic reconstruction.    
To optimize data acquisition time, we employed a non-uniform sampling across the scanning plane, with the smallest sampling period of 20 $\mu m$ concentrated around the injection position.

In principle, due to rotational symmetry, reconstructing the jet's cross-section density distribution for each height (X-position in the laboratory frame as shown in Fig. \ref{SetUp_Lebel}(a)), requires only one scan along the Y-axis (projection measurement) followed by the application of the inverse Abel transform for tomographic reconstruction.

However, to improve the SNR, which is primarily limited by beam-pointing stability, we repeated the ionization projection measurement 10 times for each cross-section, creating a ``basic projection set.''  To simulate rotational scanning, we randomly selected a projection measurement from this basic set for each virtual rotational angle before applying the inverse Radon transform for reconstruction. This method allows us to optimize acquisition time (by choosing the number of ionization projection measurements) for defined SNR requirements and specific experimental conditions.


Notably, we observed a delay of 330 $\mu \textrm{s}$ in the onset of gas expansion relative to the timing of the driving voltage pulse. This delay can be attributed to the inertia of the spring-loaded armature, which requires time to  respond to the magnetic actuation from the coil.
%
%

Following the ionizing laser pulse, various recombination processes may take place within the ionized plasma. These include three-body recombination and radiative recombination, among others \cite{Raizer_Gas_Discharge_Physics_Book_1991}.
Additionally, space charge effects such as Debye shielding \cite{Fitzpatrick_Plasma_Physics_2022} and ambipolar diffusion \cite{Raizer_Gas_Discharge_Physics_Book_1991} complicate the dynamics by affecting the movement and separation of ions and electrons within the laser-induced plasma that is exposed to the electric field of an ion detector. These phenomena can reduce the number of ions and electrons that reach the detector plates, consequently lowering the measured ion signal.
Considering the interplay of these processes, one might expect a significant, non-linear effect of gas density on the measured ion signal. However, our recent in situ strong field ionization experiments \cite{Shlomo_In_situ_arXiv_2024} using Parker's valve clearly show that this is not the case for the relevant gas densities.

A complete spatiotemporal reconstruction of gas density yields a wealth of useful information. However, focusing on specific regions of this data can be particularly insightful.

\begin{figure}
    \begin{center}
	
\includegraphics[width=\linewidth]{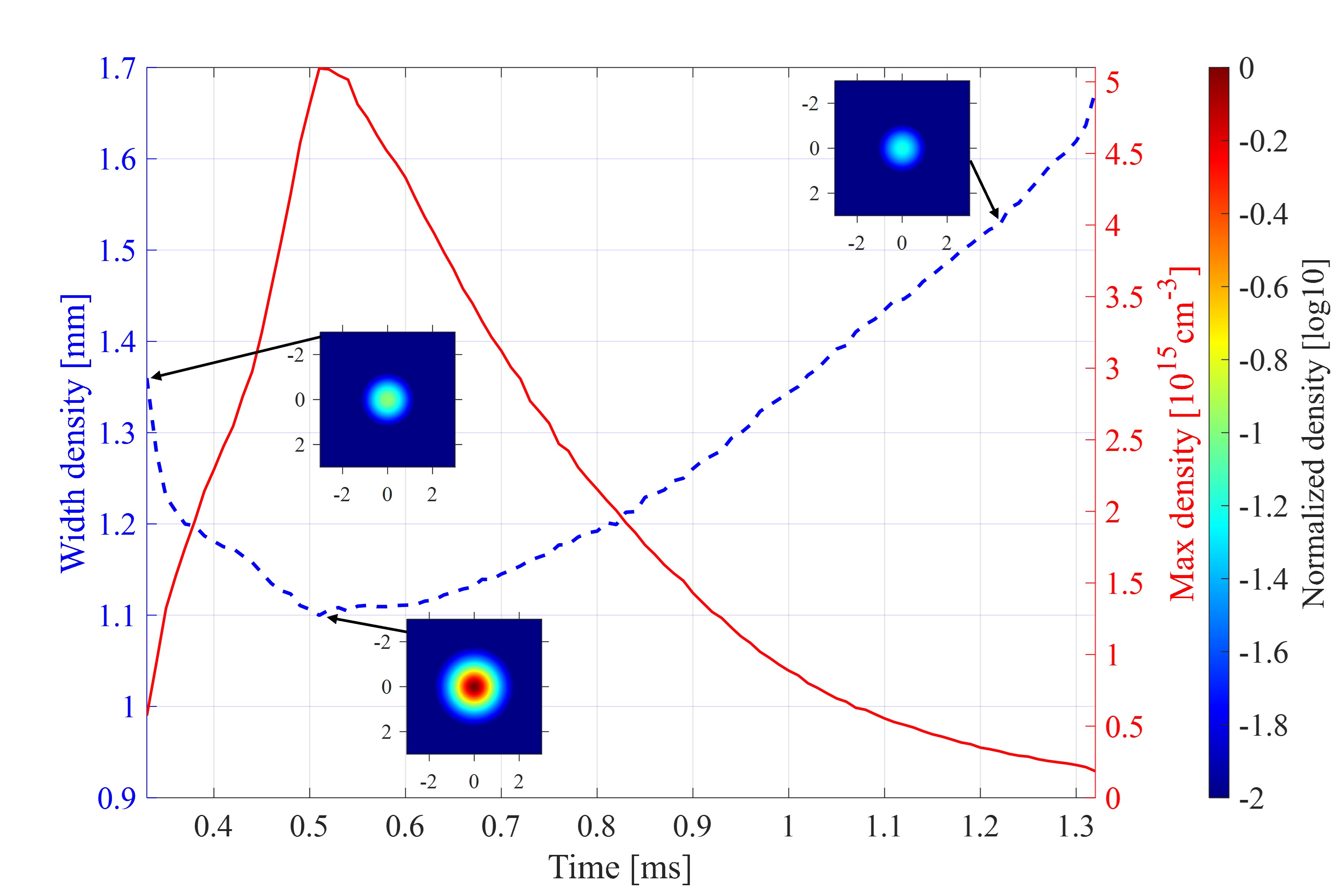}

    \caption{ \label{Slice_dynamics_vs_signal_label} Red solid curve shows the temporal evolution of the tomographically reconstructed gas density from the measured ion signal along the injection axis of the supersonic pulsed gas jet at 600 $\mu m$ from the injection point. The experiment was conducted at a back pressure of 7.5 bar. The blue dashed curve represents the evolution of the gas jet density width (measured as the second moment) in the plane perpendicular to the injection axis. The inserts depict the gas density distribution in this plane on a logarithmic scale.}
  
    \end{center}
\end{figure}

The red solid curve in Fig. \ref{Slice_dynamics_vs_signal_label} shows the temporal evolution of the  tomographically reconstructed gas density at the center of the jet, situated 600 $\mu m$ from the injection point. 
The full width at half maximum (FWHM) duration of this "density pulse" is 330 $\mu \textrm{s}$. 
Compared to our setup's time resolution of approximately 25 ps, this implies that we potentially have on the order of ten million resolution points within the time window capturing the gas expansion dynamics.

The leading edge of the argon gas jet's density profile exhibits a fast rise time of approximately $\sim160$ $\mu \textrm{s}$ ($10\%–90\%$ rise), more than three times faster than the trailing edge's fall time of  540 $\mu \textrm{s}$. 
The maximum, achieved at 520 $\mu$s, corresponds to the point with maximal ``mass'' (defined as an integral of the density over space), as depicted in the respective snapshot in Fig. \ref{Fig_Time_Evol_XZtomography}.

The blue dashed curve in Fig. \ref{Slice_dynamics_vs_signal_label} illustrates  the time evolution of the argon jet's cross-sectional width at the same 600 $\mu m$ distance. Several corresponding cross-sections of the jet at different time points are depicted in the inserts. See Supplemental Material at [URL will be inserted by publisher \url{https://drive.google.com/file/d/1rFHDyKMGM3CCnlaJf3pcRQi7IpvwTVlE/view?usp=drive_link}] for the movie illustrating comprehensive time evolution of the gas density cross-section.
A clear correlation was observed between the temporal evolution of gas density and the cross-sectional width (or equivalently, divergence angle) of the gas jet. Higher gas density correlates with a more focused (less divergent) supersonic jet. 

\begin{figure}
    \begin{center}
	
\includegraphics[width=\linewidth]{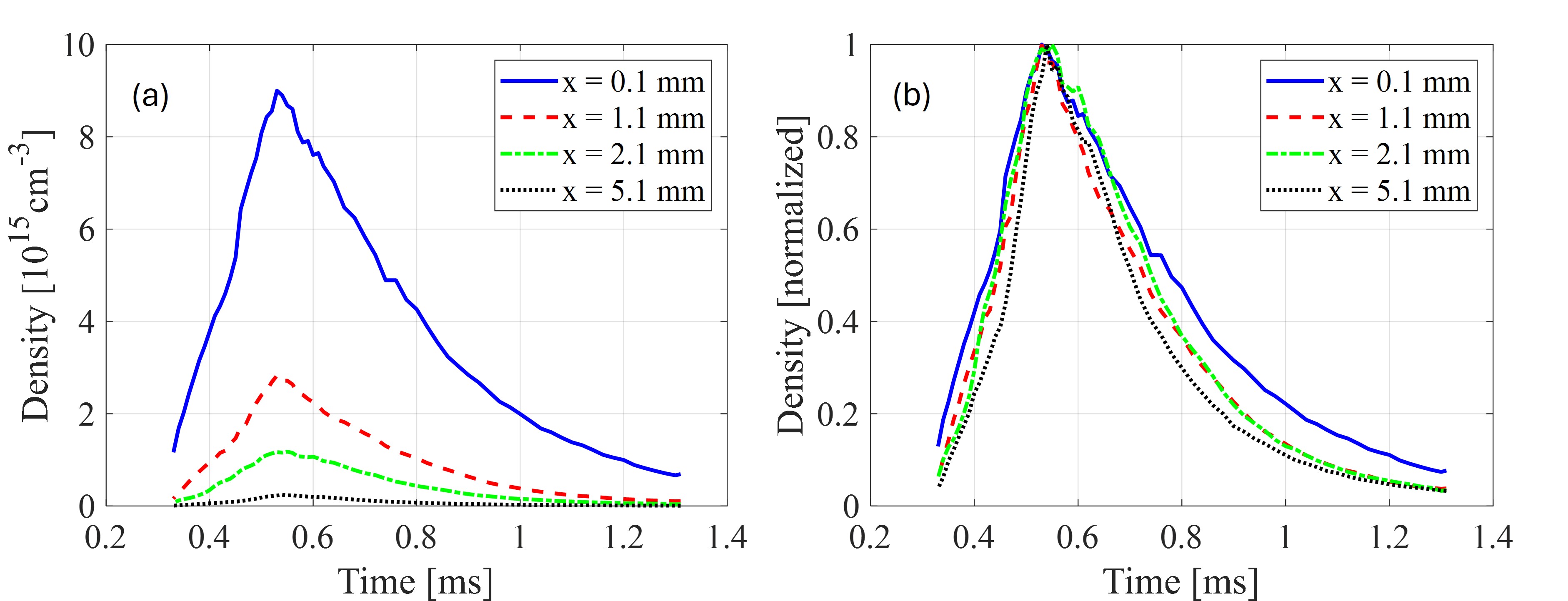}


    \caption{ \label{Fig_Ion_Signal_center_line}  The temporal evolution of the absolute (a) and normalized (b) gas density measured along the centerline of the supersonic pulsed gas jet. Experiment was conducted at 7.5 bar back pressure.}

    \end{center}
\end{figure}

 While the maximum gas density falls rapidly with the distance from the jet (Fig. \ref{Fig_Ion_Signal_center_line} (a)), as it expands in the transverse dimensions, the normalized temporal profile of the gas density along the injection axis remains quite robust, varying only slightly in an asymmetric fashion, mostly experienced by the trailing edge of the pulse, as shown in Fig. \ref{Fig_Ion_Signal_center_line} (b). 
 Surprisingly, the gas pulse duration decreases as it travels downstream.  At a distance of 100 $\mu m$, the FWHM pulse duration is 340 $\mu \textrm{s}$, while at a distance of 5.1 mm, the FWHM pulse duration decreases to 230 $\mu \textrm{s}$.

  

   The inherent fluctuations in gas density around its mean flow value \cite{Tchulov_LaserStrongFieldTom_ScRep_2017}. and the rapid transit time of the peak density in our experiment (microseconds) make it difficult to reliably determine the time delay for the peak along the jet axis. 
   This is because the inherent uncertainty in detecting the peak timing scales inversely with pulse duration and fluctuations \cite{krausz_atto_physics_Rev_Mod_Phys_2009}. However, for shorter duration pulsed jets, our approach will facilitate density pulse velocity measurements due to its inherently high temporal and spatial resolution.
   
   
  The peak moves much faster than the leading edge of the gas pulse.
   For instance, we measured a travel time of approximately 50 $\mu \textrm{s}$ for the leading edge (defined arbitrarily as 1/4 of the peak density) to traverse 5mm (from 0.1mm to 5.1mm) along the jet axis.

  Furthermore, the gas density in the trailing edge of the normalized pulse decreases faster downstream compared to the vicinity of the nozzle, likely due to its increasing expansion in the transverse directions.
   The interplay of these phenomena leads to the observed ``compression'' of the normalized gas pulse as it travels downstream.

 In addition, a characteristic "knee-like" feature is observed in the leading edge of temporal profile at approximately 430 $\mu sec$, signifying an increase in the density growth rate. 
Interestingly, this feature persists regardless of the back pressure (1.5, 4.5 and 7.5 bars) applied.
  
  
  

In conclusion, leveraging  the physics of ultrafast laser-induced strong-field ionization phenomena, we have successfully demonstrated a comprehensive 4D measurement approach, encompassing both 3D spatial dimensions and 1D temporal dimension, for gas density space-time analysis. 
Our methodology facilitates systematic quantitative measurements with exceptional spatial and temporal resolutions, suitable for both basic and applied research in far-from-equilibrium systems, including supersonic and hypersonic pulsed gas jets. 
By providing high-fidelity data, this approach will establishes a robust foundation for theoretical advancements in the study of such complex systems.

Furthermore, the presented approach further extends the utility of in situ ionization detection, rendering it highly useful and versatile tool in attosecond science and beyond. It now enables not only accurate in situ characterization of strong-field femtosecond pulses within the interaction volume \cite{Shlomo_In_situ_arXiv_2024}, but also the complete spatiotemporal characterization of the interaction medium.

We thank Michael Gedalin, Moshe Schechter, Uri Keshet and  Yuri Lyubarsky for stimulating discussions. 

Acknowledgements: E.F. acknowledges the support of Israel Science Foundation (ISF) grant 2855/21. N.S. acknowledges financial support from the Kreitman Hi-tech student fellowship.

\clearpage 
\bibliography{G:/NRC_work_since_December2007/PostDoc_Papers/Orientation_nature/Orientation_Science/Atto_references}

\begin{thebibliography}{35}%
\makeatletter
\providecommand \@ifxundefined [1]{%
 \@ifx{#1\undefined}
}%
\providecommand \@ifnum [1]{%
 \ifnum #1\expandafter \@firstoftwo
 \else \expandafter \@secondoftwo
 \fi
}%
\providecommand \@ifx [1]{%
 \ifx #1\expandafter \@firstoftwo
 \else \expandafter \@secondoftwo
 \fi
}%
\providecommand \natexlab [1]{#1}%
\providecommand \enquote  [1]{``#1''}%
\providecommand \bibnamefont  [1]{#1}%
\providecommand \bibfnamefont [1]{#1}%
\providecommand \citenamefont [1]{#1}%
\providecommand \href@noop [0]{\@secondoftwo}%
\providecommand \href [0]{\begingroup \@sanitize@url \@href}%
\providecommand \@href[1]{\@@startlink{#1}\@@href}%
\providecommand \@@href[1]{\endgroup#1\@@endlink}%
\providecommand \@sanitize@url [0]{\catcode `\\12\catcode `\$12\catcode
  `\&12\catcode `\#12\catcode `\^12\catcode `\_12\catcode `\%12\relax}%
\providecommand \@@startlink[1]{}%
\providecommand \@@endlink[0]{}%
\providecommand \url  [0]{\begingroup\@sanitize@url \@url }%
\providecommand \@url [1]{\endgroup\@href {#1}{\urlprefix }}%
\providecommand \urlprefix  [0]{URL }%
\providecommand \Eprint [0]{\href }%
\providecommand \doibase [0]{https://doi.org/}%
\providecommand \selectlanguage [0]{\@gobble}%
\providecommand \bibinfo  [0]{\@secondoftwo}%
\providecommand \bibfield  [0]{\@secondoftwo}%
\providecommand \translation [1]{[#1]}%
\providecommand \BibitemOpen [0]{}%
\providecommand \bibitemStop [0]{}%
\providecommand \bibitemNoStop [0]{.\EOS\space}%
\providecommand \EOS [0]{\spacefactor3000\relax}%
\providecommand \BibitemShut  [1]{\csname bibitem#1\endcsname}%
\let\auto@bib@innerbib\@empty
\bibitem [{\citenamefont {Modena}\ \emph {et~al.}(1995)\citenamefont {Modena},
  \citenamefont {Najmudin}, \citenamefont {Dangor}, \citenamefont {Clayton},
  \citenamefont {Marsh}, \citenamefont {Joshi}, \citenamefont {Malka},
  \citenamefont {Darrow}, \citenamefont {Danson}, \citenamefont {Neely},\ and\
  \citenamefont {Walsh}}]{Modena_Electron_nature_1995}%
  \BibitemOpen
  \bibfield  {author} {\bibinfo {author} {\bibfnamefont {A.}~\bibnamefont
  {Modena}}, \bibinfo {author} {\bibfnamefont {Z.}~\bibnamefont {Najmudin}},
  \bibinfo {author} {\bibfnamefont {A.}~\bibnamefont {Dangor}}, \bibinfo
  {author} {\bibfnamefont {C.}~\bibnamefont {Clayton}}, \bibinfo {author}
  {\bibfnamefont {K.}~\bibnamefont {Marsh}}, \bibinfo {author} {\bibfnamefont
  {C.}~\bibnamefont {Joshi}}, \bibinfo {author} {\bibfnamefont
  {V.}~\bibnamefont {Malka}}, \bibinfo {author} {\bibfnamefont
  {C.}~\bibnamefont {Darrow}}, \bibinfo {author} {\bibfnamefont
  {C.}~\bibnamefont {Danson}}, \bibinfo {author} {\bibfnamefont
  {D.}~\bibnamefont {Neely}},\ and\ \bibinfo {author} {\bibfnamefont
  {F.}~\bibnamefont {Walsh}},\ }\bibfield  {title} {\bibinfo {title} {Electron
  acceleration from the breaking of relativistic plasma waves},\ }\href@noop {}
  {\bibfield  {journal} {\bibinfo  {journal} {Nature}\ }\textbf {\bibinfo
  {volume} {377}},\ \bibinfo {pages} {606} (\bibinfo {year}
  {1995})}\BibitemShut {NoStop}%
\bibitem [{\citenamefont {Itatani}\ \emph {et~al.}(2004)\citenamefont
  {Itatani}, \citenamefont {Levesque}, \citenamefont {Zeidler}, \citenamefont
  {Niikura}, \citenamefont {P{\'e}pin}, \citenamefont {Kieffer}, \citenamefont
  {Corkum},\ and\ \citenamefont {Villeneuve}}]{Itatani_tomog_Nature2004}%
  \BibitemOpen
  \bibfield  {author} {\bibinfo {author} {\bibfnamefont {J.}~\bibnamefont
  {Itatani}}, \bibinfo {author} {\bibfnamefont {J.}~\bibnamefont {Levesque}},
  \bibinfo {author} {\bibfnamefont {D.}~\bibnamefont {Zeidler}}, \bibinfo
  {author} {\bibfnamefont {H.}~\bibnamefont {Niikura}}, \bibinfo {author}
  {\bibfnamefont {H.}~\bibnamefont {P{\'e}pin}}, \bibinfo {author}
  {\bibfnamefont {J.}~\bibnamefont {Kieffer}}, \bibinfo {author} {\bibfnamefont
  {P.}~\bibnamefont {Corkum}},\ and\ \bibinfo {author} {\bibfnamefont
  {D.}~\bibnamefont {Villeneuve}},\ }\bibfield  {title} {\bibinfo {title}
  {Tomographic imaging of molecular orbitals},\ }\href@noop {} {\bibfield
  {journal} {\bibinfo  {journal} {Nature}\ }\textbf {\bibinfo {volume} {432}},\
  \bibinfo {pages} {867} (\bibinfo {year} {2004})}\BibitemShut {NoStop}%
\bibitem [{\citenamefont {Kienberger}\ \emph {et~al.}(2004)\citenamefont
  {Kienberger}, \citenamefont {Goulielmakis}, \citenamefont {Uiberacker},
  \citenamefont {Baltuska}, \citenamefont {Yakovlev}, \citenamefont {Bammer},
  \citenamefont {Scrinzi}, \citenamefont {Westerwalbesloh}, \citenamefont
  {Kleineberg}, \citenamefont {Heinzmann},\ and\ \citenamefont
  {Krausz}}]{Kienberger_Atomic_transient_recorder_2004}%
  \BibitemOpen
  \bibfield  {author} {\bibinfo {author} {\bibfnamefont {R.}~\bibnamefont
  {Kienberger}}, \bibinfo {author} {\bibfnamefont {E.}~\bibnamefont
  {Goulielmakis}}, \bibinfo {author} {\bibfnamefont {M.}~\bibnamefont
  {Uiberacker}}, \bibinfo {author} {\bibfnamefont {A.}~\bibnamefont
  {Baltuska}}, \bibinfo {author} {\bibfnamefont {V.}~\bibnamefont {Yakovlev}},
  \bibinfo {author} {\bibfnamefont {F.}~\bibnamefont {Bammer}}, \bibinfo
  {author} {\bibfnamefont {A.}~\bibnamefont {Scrinzi}}, \bibinfo {author}
  {\bibfnamefont {T.}~\bibnamefont {Westerwalbesloh}}, \bibinfo {author}
  {\bibfnamefont {U.}~\bibnamefont {Kleineberg}}, \bibinfo {author}
  {\bibfnamefont {U.}~\bibnamefont {Heinzmann}},\ and\ \bibinfo {author}
  {\bibfnamefont {F.}~\bibnamefont {Krausz}},\ }\bibfield  {title} {\bibinfo
  {title} {Atomic transient recorder},\ }\href@noop {} {\bibfield  {journal}
  {\bibinfo  {journal} {Nature}\ }\textbf {\bibinfo {volume} {427}},\ \bibinfo
  {pages} {817} (\bibinfo {year} {2004})}\BibitemShut {NoStop}%
\bibitem [{\citenamefont {Li}\ \emph {et~al.}(2008)\citenamefont {Li},
  \citenamefont {Zhou}, \citenamefont {Lock}, \citenamefont {Patchkovskii},
  \citenamefont {Stolow}, \citenamefont {Kapteyn},\ and\ \citenamefont
  {Murnane}}]{Li_dynamics_N2O4_Science_2008}%
  \BibitemOpen
  \bibfield  {author} {\bibinfo {author} {\bibfnamefont {W.}~\bibnamefont
  {Li}}, \bibinfo {author} {\bibfnamefont {X.}~\bibnamefont {Zhou}}, \bibinfo
  {author} {\bibfnamefont {R.}~\bibnamefont {Lock}}, \bibinfo {author}
  {\bibfnamefont {S.}~\bibnamefont {Patchkovskii}}, \bibinfo {author}
  {\bibfnamefont {A.}~\bibnamefont {Stolow}}, \bibinfo {author} {\bibfnamefont
  {H.~C.}\ \bibnamefont {Kapteyn}},\ and\ \bibinfo {author} {\bibfnamefont
  {M.~M.}\ \bibnamefont {Murnane}},\ }\bibfield  {title} {\bibinfo {title}
  {Time-resolved dynamics in n(2)o(4) probed using high harmonic generation},\
  }\href {https://doi.org/10.1126/science.1163077} {\bibfield  {journal}
  {\bibinfo  {journal} {Science}\ }\textbf {\bibinfo {volume} {322}},\ \bibinfo
  {pages} {1207} (\bibinfo {year} {2008})}\BibitemShut {NoStop}%
\bibitem [{\citenamefont {Denavit}\ and\ \citenamefont
  {Phillion}(1994)}]{Denavit__gas_targets_PhysPlasmas_1994}%
  \BibitemOpen
  \bibfield  {author} {\bibinfo {author} {\bibfnamefont {J.}~\bibnamefont
  {Denavit}}\ and\ \bibinfo {author} {\bibfnamefont {D.}~\bibnamefont
  {Phillion}},\ }\bibfield  {title} {\bibinfo {title} {Laser ionization and
  heating of gas targets for long-scale-length instability experiments},\
  }\href@noop {} {\bibfield  {journal} {\bibinfo  {journal} {Phys. Plasmas}\
  }\textbf {\bibinfo {volume} {1}},\ \bibinfo {pages} {1971} (\bibinfo {year}
  {1994})}\BibitemShut {NoStop}%
\bibitem [{\citenamefont {Failor}\ \emph {et~al.}(2003)\citenamefont {Failor},
  \citenamefont {Chantrenne}, \citenamefont {Coleman}, \citenamefont {Levine},
  \citenamefont {Song},\ and\ \citenamefont
  {Sze}}]{Failor_fluorescence_gas_tomogr_RSI_2003}%
  \BibitemOpen
  \bibfield  {author} {\bibinfo {author} {\bibfnamefont {B.}~\bibnamefont
  {Failor}}, \bibinfo {author} {\bibfnamefont {S.}~\bibnamefont {Chantrenne}},
  \bibinfo {author} {\bibfnamefont {P.}~\bibnamefont {Coleman}}, \bibinfo
  {author} {\bibfnamefont {J.}~\bibnamefont {Levine}}, \bibinfo {author}
  {\bibfnamefont {Y.}~\bibnamefont {Song}},\ and\ \bibinfo {author}
  {\bibfnamefont {H.}~\bibnamefont {Sze}},\ }\bibfield  {title} {\bibinfo
  {title} {Proof-of-principle laser-induced fluorescence measurements of gas
  distributions from supersonic nozzles},\ }\href@noop {} {\bibfield  {journal}
  {\bibinfo  {journal} {Rev. Sci. Instrum.}\ }\textbf {\bibinfo {volume}
  {74}},\ \bibinfo {pages} {1070} (\bibinfo {year} {2003})}\BibitemShut
  {NoStop}%
\bibitem [{\citenamefont {Thorpe}\ \emph {et~al.}(2009)\citenamefont {Thorpe},
  \citenamefont {Adler}, \citenamefont {Cossel}, \citenamefont {de~Miranda},\
  and\ \citenamefont {Ye}}]{Thorpe_tomography_CPL_2009}%
  \BibitemOpen
  \bibfield  {author} {\bibinfo {author} {\bibfnamefont {M.}~\bibnamefont
  {Thorpe}}, \bibinfo {author} {\bibfnamefont {F.}~\bibnamefont {Adler}},
  \bibinfo {author} {\bibfnamefont {K.}~\bibnamefont {Cossel}}, \bibinfo
  {author} {\bibfnamefont {M.}~\bibnamefont {de~Miranda}},\ and\ \bibinfo
  {author} {\bibfnamefont {J.}~\bibnamefont {Ye}},\ }\bibfield  {title}
  {\bibinfo {title} {Tomography of a supersonically cooled molecular jet using
  cavity-enhanced direct frequency comb spectroscopy},\ }\href@noop {}
  {\bibfield  {journal} {\bibinfo  {journal} {Chem. Phys. Lett.}\ }\textbf
  {\bibinfo {volume} {468}},\ \bibinfo {pages} {1} (\bibinfo {year}
  {2009})}\BibitemShut {NoStop}%
\bibitem [{\citenamefont {Henson}\ \emph {et~al.}(2012)\citenamefont {Henson},
  \citenamefont {Gersten}, \citenamefont {Shagam}, \citenamefont {Narevicius},\
  and\ \citenamefont
  {Narevicius}}]{Henson_Observation_resonances_beams_Science_2012}%
  \BibitemOpen
  \bibfield  {author} {\bibinfo {author} {\bibfnamefont {A.}~\bibnamefont
  {Henson}}, \bibinfo {author} {\bibfnamefont {S.}~\bibnamefont {Gersten}},
  \bibinfo {author} {\bibfnamefont {Y.}~\bibnamefont {Shagam}}, \bibinfo
  {author} {\bibfnamefont {J.}~\bibnamefont {Narevicius}},\ and\ \bibinfo
  {author} {\bibfnamefont {E.}~\bibnamefont {Narevicius}},\ }\bibfield  {title}
  {\bibinfo {title} {Observation of resonances in penning ionization reactions
  at sub-kelvin temperatures in merged beams},\ }\href@noop {} {\bibfield
  {journal} {\bibinfo  {journal} {Science}\ }\textbf {\bibinfo {volume}
  {338}},\ \bibinfo {pages} {234} (\bibinfo {year} {2012})}\BibitemShut
  {NoStop}%
\bibitem [{\citenamefont {Bellini}\ \emph {et~al.}(1998)\citenamefont
  {Bellini}, \citenamefont {Lynga}, \citenamefont {Tozzi}, \citenamefont
  {Gaarde}, \citenamefont {H{\"a}nsch}, \citenamefont {L'Huillier},\ and\
  \citenamefont {Wahlstr{\"o}m}}]{Bellini_temp_coherence_PRL1998}%
  \BibitemOpen
  \bibfield  {author} {\bibinfo {author} {\bibfnamefont {M.}~\bibnamefont
  {Bellini}}, \bibinfo {author} {\bibfnamefont {C.}~\bibnamefont {Lynga}},
  \bibinfo {author} {\bibfnamefont {A.}~\bibnamefont {Tozzi}}, \bibinfo
  {author} {\bibfnamefont {M.~B.}\ \bibnamefont {Gaarde}}, \bibinfo {author}
  {\bibfnamefont {T.~W.}\ \bibnamefont {H{\"a}nsch}}, \bibinfo {author}
  {\bibfnamefont {A.}~\bibnamefont {L'Huillier}},\ and\ \bibinfo {author}
  {\bibfnamefont {C.~G.}\ \bibnamefont {Wahlstr{\"o}m}},\ }\bibfield  {title}
  {\bibinfo {title} {Temporal coherence of ultrashort high-order harmonic
  pulses},\ }\href@noop {} {\bibfield  {journal} {\bibinfo  {journal} {Phys.
  Rev. Lett.}\ }\textbf {\bibinfo {volume} {81}},\ \bibinfo {pages} {297}
  (\bibinfo {year} {1998})}\BibitemShut {NoStop}%
\bibitem [{\citenamefont {Frumker}\ \emph
  {et~al.}(2012{\natexlab{a}})\citenamefont {Frumker}, \citenamefont {Kajumba},
  \citenamefont {Bertrand}, \citenamefont {W{\"o}rner}, \citenamefont
  {Hebeisen}, \citenamefont {Hockett}, \citenamefont {Spanner}, \citenamefont
  {Patchkovskii}, \citenamefont {Paulus}, \citenamefont {Villeneuve},\ and\
  \citenamefont {Corkum}}]{Frumker_orientation_HHG_PRL2012}%
  \BibitemOpen
  \bibfield  {author} {\bibinfo {author} {\bibfnamefont {E.}~\bibnamefont
  {Frumker}}, \bibinfo {author} {\bibfnamefont {N.}~\bibnamefont {Kajumba}},
  \bibinfo {author} {\bibfnamefont {J.~B.}\ \bibnamefont {Bertrand}}, \bibinfo
  {author} {\bibfnamefont {H.~J.}\ \bibnamefont {W{\"o}rner}}, \bibinfo
  {author} {\bibfnamefont {C.~T.}\ \bibnamefont {Hebeisen}}, \bibinfo {author}
  {\bibfnamefont {P.}~\bibnamefont {Hockett}}, \bibinfo {author} {\bibfnamefont
  {M.}~\bibnamefont {Spanner}}, \bibinfo {author} {\bibfnamefont
  {S.}~\bibnamefont {Patchkovskii}}, \bibinfo {author} {\bibfnamefont {G.~G.}\
  \bibnamefont {Paulus}}, \bibinfo {author} {\bibfnamefont {D.~M.}\
  \bibnamefont {Villeneuve}},\ and\ \bibinfo {author} {\bibfnamefont {P.~B.}\
  \bibnamefont {Corkum}},\ }\bibfield  {title} {\bibinfo {title} {Probing polar
  molecules with high harmonic spectroscopy},\ }\href@noop {} {\bibfield
  {journal} {\bibinfo  {journal} {Phys. Rev. Lett.}\ }\textbf {\bibinfo
  {volume} {109}},\ \bibinfo {pages} {233904} (\bibinfo {year}
  {2012}{\natexlab{a}})}\BibitemShut {NoStop}%
\bibitem [{\citenamefont {Frumker}\ \emph
  {et~al.}(2012{\natexlab{b}})\citenamefont {Frumker}, \citenamefont
  {Hebeisen}, \citenamefont {Kajumba}, \citenamefont {Bertrand}, \citenamefont
  {W{\"o}rner}, \citenamefont {Spanner}, \citenamefont {Villeneuve},
  \citenamefont {Naumov},\ and\ \citenamefont
  {Corkum}}]{frumker_wavepacket_PRL2012}%
  \BibitemOpen
  \bibfield  {author} {\bibinfo {author} {\bibfnamefont {E.}~\bibnamefont
  {Frumker}}, \bibinfo {author} {\bibfnamefont {C.}~\bibnamefont {Hebeisen}},
  \bibinfo {author} {\bibfnamefont {N.}~\bibnamefont {Kajumba}}, \bibinfo
  {author} {\bibfnamefont {J.}~\bibnamefont {Bertrand}}, \bibinfo {author}
  {\bibfnamefont {H.}~\bibnamefont {W{\"o}rner}}, \bibinfo {author}
  {\bibfnamefont {M.}~\bibnamefont {Spanner}}, \bibinfo {author} {\bibfnamefont
  {D.}~\bibnamefont {Villeneuve}}, \bibinfo {author} {\bibfnamefont
  {A.}~\bibnamefont {Naumov}},\ and\ \bibinfo {author} {\bibfnamefont
  {P.}~\bibnamefont {Corkum}},\ }\bibfield  {title} {\bibinfo {title} {Oriented
  rotational wave-packet dynamics studies via high harmonic generation},\
  }\href@noop {} {\bibfield  {journal} {\bibinfo  {journal} {Phys. Rev. Lett.}\
  }\textbf {\bibinfo {volume} {109}},\ \bibinfo {pages} {113901} (\bibinfo
  {year} {2012}{\natexlab{b}})}\BibitemShut {NoStop}%
\bibitem [{\citenamefont {Even}\ \emph {et~al.}(2000)\citenamefont {Even},
  \citenamefont {Jortner}, \citenamefont {Noy}, \citenamefont {Lavie},\ and\
  \citenamefont {Cossart-Magos}}]{even_cooling1K_JCP_2000}%
  \BibitemOpen
  \bibfield  {author} {\bibinfo {author} {\bibfnamefont {U.}~\bibnamefont
  {Even}}, \bibinfo {author} {\bibfnamefont {J.}~\bibnamefont {Jortner}},
  \bibinfo {author} {\bibfnamefont {D.}~\bibnamefont {Noy}}, \bibinfo {author}
  {\bibfnamefont {N.}~\bibnamefont {Lavie}},\ and\ \bibinfo {author}
  {\bibfnamefont {C.}~\bibnamefont {Cossart-Magos}},\ }\bibfield  {title}
  {\bibinfo {title} {Cooling of large molecules below 1 k and he clusters
  formation},\ }\href@noop {} {\bibfield  {journal} {\bibinfo  {journal} {The
  J. of Chem. Phys.}\ }\textbf {\bibinfo {volume} {112}},\ \bibinfo {pages}
  {8068} (\bibinfo {year} {2000})}\BibitemShut {NoStop}%
\bibitem [{\citenamefont {Narevicius}\ \emph {et~al.}(2008)\citenamefont
  {Narevicius}, \citenamefont {Libson}, \citenamefont {Parthey}, \citenamefont
  {Chavez}, \citenamefont {Narevicius}, \citenamefont {Even},\ and\
  \citenamefont {Raizen}}]{Narevicius_Stopping_beams_PRL_2008}%
  \BibitemOpen
  \bibfield  {author} {\bibinfo {author} {\bibfnamefont {E.}~\bibnamefont
  {Narevicius}}, \bibinfo {author} {\bibfnamefont {A.}~\bibnamefont {Libson}},
  \bibinfo {author} {\bibfnamefont {C.}~\bibnamefont {Parthey}}, \bibinfo
  {author} {\bibfnamefont {I.}~\bibnamefont {Chavez}}, \bibinfo {author}
  {\bibfnamefont {J.}~\bibnamefont {Narevicius}}, \bibinfo {author}
  {\bibfnamefont {U.}~\bibnamefont {Even}},\ and\ \bibinfo {author}
  {\bibfnamefont {M.}~\bibnamefont {Raizen}},\ }\bibfield  {title} {\bibinfo
  {title} {Stopping supersonic beams with a series of pulsed electromagnetic
  coils: an atomic coilgun},\ }\href@noop {} {\bibfield  {journal} {\bibinfo
  {journal} {Phys. Rev. Lett.}\ }\textbf {\bibinfo {volume} {100}},\ \bibinfo
  {pages} {093003} (\bibinfo {year} {2008})}\BibitemShut {NoStop}%
\bibitem [{\citenamefont {Malka}\ \emph {et~al.}(2000)\citenamefont {Malka},
  \citenamefont {Coulaud}, \citenamefont {Geindre}, \citenamefont {Lopez},
  \citenamefont {Najmudin}, \citenamefont {Neely},\ and\ \citenamefont
  {Amiranoff}}]{malka2000charact}%
  \BibitemOpen
  \bibfield  {author} {\bibinfo {author} {\bibfnamefont {V.}~\bibnamefont
  {Malka}}, \bibinfo {author} {\bibfnamefont {C.}~\bibnamefont {Coulaud}},
  \bibinfo {author} {\bibfnamefont {J.}~\bibnamefont {Geindre}}, \bibinfo
  {author} {\bibfnamefont {V.}~\bibnamefont {Lopez}}, \bibinfo {author}
  {\bibfnamefont {Z.}~\bibnamefont {Najmudin}}, \bibinfo {author}
  {\bibfnamefont {D.}~\bibnamefont {Neely}},\ and\ \bibinfo {author}
  {\bibfnamefont {F.}~\bibnamefont {Amiranoff}},\ }\bibfield  {title} {\bibinfo
  {title} {Characterization of neutral density profile in a wide range of
  pressure of cylindrical pulsed gas jets},\ }\href@noop {} {\bibfield
  {journal} {\bibinfo  {journal} {Rev. Sci. Instrum.}\ }\textbf {\bibinfo
  {volume} {71}},\ \bibinfo {pages} {2329} (\bibinfo {year}
  {2000})}\BibitemShut {NoStop}%
\bibitem [{\citenamefont {Schofield}\ \emph {et~al.}(2009)\citenamefont
  {Schofield}, \citenamefont {Paganin},\ and\ \citenamefont
  {Bishop}}]{Sschofield_Absolute_density_tomography_RSI_2009}%
  \BibitemOpen
  \bibfield  {author} {\bibinfo {author} {\bibfnamefont {N.}~\bibnamefont
  {Schofield}}, \bibinfo {author} {\bibfnamefont {D.}~\bibnamefont {Paganin}},\
  and\ \bibinfo {author} {\bibfnamefont {A.}~\bibnamefont {Bishop}},\
  }\bibfield  {title} {\bibinfo {title} {Absolute density-profile tomography of
  molecular beams using multiphoton ionization},\ }\href@noop {} {\bibfield
  {journal} {\bibinfo  {journal} {Rev. Sci. Instrum.}\ }\textbf {\bibinfo
  {volume} {80}},\ \bibinfo {pages} {123105} (\bibinfo {year}
  {2009})}\BibitemShut {NoStop}%
\bibitem [{\citenamefont {Wachulak}\ \emph {et~al.}(2014)\citenamefont
  {Wachulak}, \citenamefont {Z{\'a}pra{\v{z}}n{\`y}}, \citenamefont {Bartnik},
  \citenamefont {Fok}, \citenamefont {Jarocki}, \citenamefont {Kostecki},
  \citenamefont {Szczurek}, \citenamefont {Koryt{\'a}r}, \citenamefont
  {Fiedorowicz} \emph {et~al.}}]{Wachulak_Extreme_UV_Tomog_AppPhysB_2014}%
  \BibitemOpen
  \bibfield  {author} {\bibinfo {author} {\bibfnamefont {P.}~\bibnamefont
  {Wachulak}}, \bibinfo {author} {\bibfnamefont {Z.}~\bibnamefont
  {Z{\'a}pra{\v{z}}n{\`y}}}, \bibinfo {author} {\bibfnamefont {A.}~\bibnamefont
  {Bartnik}}, \bibinfo {author} {\bibfnamefont {T.}~\bibnamefont {Fok}},
  \bibinfo {author} {\bibfnamefont {R.}~\bibnamefont {Jarocki}}, \bibinfo
  {author} {\bibfnamefont {J.}~\bibnamefont {Kostecki}}, \bibinfo {author}
  {\bibfnamefont {M.}~\bibnamefont {Szczurek}}, \bibinfo {author}
  {\bibfnamefont {D.}~\bibnamefont {Koryt{\'a}r}}, \bibinfo {author}
  {\bibfnamefont {H.}~\bibnamefont {Fiedorowicz}}, \emph {et~al.},\ }\bibfield
  {title} {\bibinfo {title} {Extreme ultraviolet tomography of multi-jet gas
  puff target for high-order harmonic generation},\ }\href@noop {} {\bibfield
  {journal} {\bibinfo  {journal} {Appl. Phys. B}\ ,\ \bibinfo {pages} {1}}
  (\bibinfo {year} {2014})}\BibitemShut {NoStop}%
\bibitem [{\citenamefont {Tejeda}\ \emph {et~al.}(1996)\citenamefont {Tejeda},
  \citenamefont {Mat{\'e}}, \citenamefont {Fern{\'a}ndez-S{\'a}nchez},\ and\
  \citenamefont {Montero}}]{Tejeda_mapping_supersonic_PRL_1996}%
  \BibitemOpen
  \bibfield  {author} {\bibinfo {author} {\bibfnamefont {G.}~\bibnamefont
  {Tejeda}}, \bibinfo {author} {\bibfnamefont {B.}~\bibnamefont {Mat{\'e}}},
  \bibinfo {author} {\bibfnamefont {J.}~\bibnamefont
  {Fern{\'a}ndez-S{\'a}nchez}},\ and\ \bibinfo {author} {\bibfnamefont
  {S.}~\bibnamefont {Montero}},\ }\bibfield  {title} {\bibinfo {title}
  {Temperature and density mapping of supersonic jet expansions using linear
  raman spectroscopy},\ }\href@noop {} {\bibfield  {journal} {\bibinfo
  {journal} {Phys. Rev. Lett.}\ }\textbf {\bibinfo {volume} {76}},\ \bibinfo
  {pages} {34} (\bibinfo {year} {1996})}\BibitemShut {NoStop}%
\bibitem [{\citenamefont {Comby}\ \emph {et~al.}(2018)\citenamefont {Comby},
  \citenamefont {Beaulieu}, \citenamefont {Constant}, \citenamefont {Descamps},
  \citenamefont {Petit},\ and\ \citenamefont
  {Mairesse}}]{Comby_Absolute_Gas_Density_OE_2018}%
  \BibitemOpen
  \bibfield  {author} {\bibinfo {author} {\bibfnamefont {A.}~\bibnamefont
  {Comby}}, \bibinfo {author} {\bibfnamefont {S.}~\bibnamefont {Beaulieu}},
  \bibinfo {author} {\bibfnamefont {E.}~\bibnamefont {Constant}}, \bibinfo
  {author} {\bibfnamefont {D.}~\bibnamefont {Descamps}}, \bibinfo {author}
  {\bibfnamefont {S.}~\bibnamefont {Petit}},\ and\ \bibinfo {author}
  {\bibfnamefont {Y.}~\bibnamefont {Mairesse}},\ }\bibfield  {title} {\bibinfo
  {title} {Absolute gas density profiling in high-order harmonic generation},\
  }\href@noop {} {\bibfield  {journal} {\bibinfo  {journal} {Optics express}\
  }\textbf {\bibinfo {volume} {26}},\ \bibinfo {pages} {6001} (\bibinfo {year}
  {2018})}\BibitemShut {NoStop}%
\bibitem [{\citenamefont {Tchulov}\ \emph {et~al.}(2017)\citenamefont
  {Tchulov}, \citenamefont {Negro}, \citenamefont {Stagira}, \citenamefont
  {Devetta}, \citenamefont {Vozzi},\ and\ \citenamefont
  {Frumker}}]{Tchulov_LaserStrongFieldTom_ScRep_2017}%
  \BibitemOpen
  \bibfield  {author} {\bibinfo {author} {\bibfnamefont {O.}~\bibnamefont
  {Tchulov}}, \bibinfo {author} {\bibfnamefont {M.}~\bibnamefont {Negro}},
  \bibinfo {author} {\bibfnamefont {S.}~\bibnamefont {Stagira}}, \bibinfo
  {author} {\bibfnamefont {M.}~\bibnamefont {Devetta}}, \bibinfo {author}
  {\bibfnamefont {C.}~\bibnamefont {Vozzi}},\ and\ \bibinfo {author}
  {\bibfnamefont {E.}~\bibnamefont {Frumker}},\ }\bibfield  {title} {\bibinfo
  {title} {Laser induced strong-field ionization gas jet tomography},\
  }\href@noop {} {\bibfield  {journal} {\bibinfo  {journal} {Sci. Rep.}\
  }\textbf {\bibinfo {volume} {7}},\ \bibinfo {pages} {1} (\bibinfo {year}
  {2017})}\BibitemShut {NoStop}%
\bibitem [{\citenamefont {Hagmeister}\ \emph {et~al.}(2022)\citenamefont
  {Hagmeister}, \citenamefont {Hemmers},\ and\ \citenamefont
  {Pretzler}}]{Hagmeister_HHG_Characterization_Jet_AppPhysB_2022}%
  \BibitemOpen
  \bibfield  {author} {\bibinfo {author} {\bibfnamefont {B.}~\bibnamefont
  {Hagmeister}}, \bibinfo {author} {\bibfnamefont {D.}~\bibnamefont
  {Hemmers}},\ and\ \bibinfo {author} {\bibfnamefont {G.}~\bibnamefont
  {Pretzler}},\ }\bibfield  {title} {\bibinfo {title} {Characterization of a
  pulsed, supersonic gas jet by absorption of high-order harmonics in the
  extreme ultraviolet spectral range},\ }\href@noop {} {\bibfield  {journal}
  {\bibinfo  {journal} {Appl. Phys. B}\ }\textbf {\bibinfo {volume} {128}},\
  \bibinfo {pages} {172} (\bibinfo {year} {2022})}\BibitemShut {NoStop}%
\bibitem [{\citenamefont {Liu}\ \emph {et~al.}(2023)\citenamefont {Liu},
  \citenamefont {Ma}, \citenamefont {Zhang}, \citenamefont {Lv}, \citenamefont
  {Song}, \citenamefont {Wang}, \citenamefont {Yang}, \citenamefont {Yang},
  \citenamefont {Wang}, \citenamefont {Li},\ and\ \citenamefont
  {Zhao}}]{Liu_SpaceTime_Jet_FiP_2023}%
  \BibitemOpen
  \bibfield  {author} {\bibinfo {author} {\bibfnamefont {Q.}~\bibnamefont
  {Liu}}, \bibinfo {author} {\bibfnamefont {M.}~\bibnamefont {Ma}}, \bibinfo
  {author} {\bibfnamefont {X.}~\bibnamefont {Zhang}}, \bibinfo {author}
  {\bibfnamefont {C.}~\bibnamefont {Lv}}, \bibinfo {author} {\bibfnamefont
  {J.}~\bibnamefont {Song}}, \bibinfo {author} {\bibfnamefont {Z.}~\bibnamefont
  {Wang}}, \bibinfo {author} {\bibfnamefont {G.}~\bibnamefont {Yang}}, \bibinfo
  {author} {\bibfnamefont {Y.}~\bibnamefont {Yang}}, \bibinfo {author}
  {\bibfnamefont {J.}~\bibnamefont {Wang}}, \bibinfo {author} {\bibfnamefont
  {Q.}~\bibnamefont {Li}},\ and\ \bibinfo {author} {\bibfnamefont
  {B.}~\bibnamefont {Zhao}},\ }\bibfield  {title} {\bibinfo {title}
  {Characteristic diagnosis of supersonic gas jet target for laser wakefield
  acceleration with high spatial-temporal resolution normaski interference
  system},\ }\href@noop {} {\bibfield  {journal} {\bibinfo  {journal} {Front.
  Phys.}\ }\textbf {\bibinfo {volume} {11}},\ \bibinfo {pages} {1203946}
  (\bibinfo {year} {2023})}\BibitemShut {NoStop}%
\bibitem [{\citenamefont {Nagyill{\'e}s}\ \emph {et~al.}(2023)\citenamefont
  {Nagyill{\'e}s}, \citenamefont {Diveki}, \citenamefont {Nayak}, \citenamefont
  {Dumergue}, \citenamefont {Major}, \citenamefont {Varj{\'u}},\ and\
  \citenamefont {Kahaly}}]{Nagyilles_time_resolved_PysRevAppl_2023}%
  \BibitemOpen
  \bibfield  {author} {\bibinfo {author} {\bibfnamefont {B.}~\bibnamefont
  {Nagyill{\'e}s}}, \bibinfo {author} {\bibfnamefont {Z.}~\bibnamefont
  {Diveki}}, \bibinfo {author} {\bibfnamefont {A.}~\bibnamefont {Nayak}},
  \bibinfo {author} {\bibfnamefont {M.}~\bibnamefont {Dumergue}}, \bibinfo
  {author} {\bibfnamefont {B.}~\bibnamefont {Major}}, \bibinfo {author}
  {\bibfnamefont {K.}~\bibnamefont {Varj{\'u}}},\ and\ \bibinfo {author}
  {\bibfnamefont {S.}~\bibnamefont {Kahaly}},\ }\bibfield  {title} {\bibinfo
  {title} {Time-resolved investigation of a high-repetition-rate gas-jet target
  for high-harmonic generation},\ }\href@noop {} {\bibfield  {journal}
  {\bibinfo  {journal} {Phys. Rev. Appl}\ }\textbf {\bibinfo {volume} {20}},\
  \bibinfo {pages} {054048} (\bibinfo {year} {2023})}\BibitemShut {NoStop}%
\bibitem [{\citenamefont {Frumker}()}]{Frumker_thory_SFI_tomography_2024}%
  \BibitemOpen
  \bibfield  {author} {\bibinfo {author} {\bibfnamefont {E.}~\bibnamefont
  {Frumker}},\ }\bibfield  {title} {\bibinfo {title} {Theory of laser induced
  strong-field ionization imaging and tomography},\ }\bibinfo {note} {(to be
  published)}\BibitemShut {NoStop}%
\bibitem [{\citenamefont {Gannaway}\ and\ \citenamefont
  {Sheppard}(1978)}]{Gannaway_Scanning_Micr_SHG_OQE_1978}%
  \BibitemOpen
  \bibfield  {author} {\bibinfo {author} {\bibfnamefont {J.}~\bibnamefont
  {Gannaway}}\ and\ \bibinfo {author} {\bibfnamefont {C.}~\bibnamefont
  {Sheppard}},\ }\bibfield  {title} {\bibinfo {title} {Second-harmonic imaging
  in the scanning optical microscope},\ }\href@noop {} {\bibfield  {journal}
  {\bibinfo  {journal} {Opt. Quantum Electron.}\ }\textbf {\bibinfo {volume}
  {10}},\ \bibinfo {pages} {435} (\bibinfo {year} {1978})}\BibitemShut
  {NoStop}%
\bibitem [{\citenamefont {Denk}\ \emph {et~al.}(1990)\citenamefont {Denk},
  \citenamefont {Strickler},\ and\ \citenamefont
  {Webb}}]{Denk_TPF_microscopy_Science_1990}%
  \BibitemOpen
  \bibfield  {author} {\bibinfo {author} {\bibfnamefont {W.}~\bibnamefont
  {Denk}}, \bibinfo {author} {\bibfnamefont {J.}~\bibnamefont {Strickler}},\
  and\ \bibinfo {author} {\bibfnamefont {W.}~\bibnamefont {Webb}},\ }\bibfield
  {title} {\bibinfo {title} {Two-photon laser scanning fluorescence
  microscopy},\ }\href@noop {} {\bibfield  {journal} {\bibinfo  {journal}
  {Science}\ }\textbf {\bibinfo {volume} {248}},\ \bibinfo {pages} {73}
  (\bibinfo {year} {1990})}\BibitemShut {NoStop}%
\bibitem [{\citenamefont {Barad}\ \emph {et~al.}(1997)\citenamefont {Barad},
  \citenamefont {Eisenberg}, \citenamefont {H.M.},\ and\ \citenamefont
  {Silberberg}}]{Barad_THG_microscopy_APL1997}%
  \BibitemOpen
  \bibfield  {author} {\bibinfo {author} {\bibfnamefont {Y.}~\bibnamefont
  {Barad}}, \bibinfo {author} {\bibfnamefont {H.}~\bibnamefont {Eisenberg}},
  \bibinfo {author} {\bibnamefont {H.M.}},\ and\ \bibinfo {author}
  {\bibfnamefont {Y.}~\bibnamefont {Silberberg}},\ }\bibfield  {title}
  {\bibinfo {title} {Nonlinear scanning laser microscopy by third harmonic
  generation},\ }\href@noop {} {\bibfield  {journal} {\bibinfo  {journal}
  {Appl. Phys. Lett.}\ }\textbf {\bibinfo {volume} {70}},\ \bibinfo {pages}
  {922} (\bibinfo {year} {1997})}\BibitemShut {NoStop}%
\bibitem [{\citenamefont {Freudiger}\ \emph {et~al.}(2008)\citenamefont
  {Freudiger}, \citenamefont {Min}, \citenamefont {Saar}, \citenamefont {Lu},
  \citenamefont {Holtom}, \citenamefont {He}, \citenamefont {Tsai},
  \citenamefont {Kang},\ and\ \citenamefont
  {Xie}}]{Freudiger_SRS_Microscopy_Science_2008}%
  \BibitemOpen
  \bibfield  {author} {\bibinfo {author} {\bibfnamefont {C.~W.}\ \bibnamefont
  {Freudiger}}, \bibinfo {author} {\bibfnamefont {W.}~\bibnamefont {Min}},
  \bibinfo {author} {\bibfnamefont {B.}~\bibnamefont {Saar}}, \bibinfo {author}
  {\bibfnamefont {S.}~\bibnamefont {Lu}}, \bibinfo {author} {\bibfnamefont
  {G.}~\bibnamefont {Holtom}}, \bibinfo {author} {\bibfnamefont
  {C.}~\bibnamefont {He}}, \bibinfo {author} {\bibfnamefont {J.}~\bibnamefont
  {Tsai}}, \bibinfo {author} {\bibfnamefont {J.}~\bibnamefont {Kang}},\ and\
  \bibinfo {author} {\bibfnamefont {X.}~\bibnamefont {Xie}},\ }\bibfield
  {title} {\bibinfo {title} {Label-free biomedical imaging with high
  sensitivity by stimulated raman scattering microscopy},\ }\href@noop {}
  {\bibfield  {journal} {\bibinfo  {journal} {Science}\ }\textbf {\bibinfo
  {volume} {322}},\ \bibinfo {pages} {1857} (\bibinfo {year}
  {2008})}\BibitemShut {NoStop}%
\bibitem [{\citenamefont {Vozzi}\ \emph {et~al.}(2011)\citenamefont {Vozzi},
  \citenamefont {Negro}, \citenamefont {Calegari}, \citenamefont {Sansone},
  \citenamefont {Nisoli}, \citenamefont {De~Silvestri},\ and\ \citenamefont
  {Stagira}}]{Vozzi_gen_tomography_NatPhys_2011}%
  \BibitemOpen
  \bibfield  {author} {\bibinfo {author} {\bibfnamefont {C.}~\bibnamefont
  {Vozzi}}, \bibinfo {author} {\bibfnamefont {M.}~\bibnamefont {Negro}},
  \bibinfo {author} {\bibfnamefont {F.}~\bibnamefont {Calegari}}, \bibinfo
  {author} {\bibfnamefont {G.}~\bibnamefont {Sansone}}, \bibinfo {author}
  {\bibfnamefont {M.}~\bibnamefont {Nisoli}}, \bibinfo {author} {\bibfnamefont
  {S.}~\bibnamefont {De~Silvestri}},\ and\ \bibinfo {author} {\bibfnamefont
  {S.}~\bibnamefont {Stagira}},\ }\bibfield  {title} {\bibinfo {title}
  {Generalized molecular orbital tomography},\ }\href@noop {} {\bibfield
  {journal} {\bibinfo  {journal} {Nat. Phys.}\ }\textbf {\bibinfo {volume}
  {7}},\ \bibinfo {pages} {823} (\bibinfo {year} {2011})}\BibitemShut {NoStop}%
\bibitem [{\citenamefont {Litvinyuk}\ \emph {et~al.}(2003)\citenamefont
  {Litvinyuk}, \citenamefont {Lee}, \citenamefont {Dooley}, \citenamefont
  {Rayner}, \citenamefont {Villeneuve},\ and\ \citenamefont
  {Corkum}}]{Litvinyuk_N2_ion_PRL_2003}%
  \BibitemOpen
  \bibfield  {author} {\bibinfo {author} {\bibfnamefont {I.~V.}\ \bibnamefont
  {Litvinyuk}}, \bibinfo {author} {\bibfnamefont {K.~F.}\ \bibnamefont {Lee}},
  \bibinfo {author} {\bibfnamefont {P.~W.}\ \bibnamefont {Dooley}}, \bibinfo
  {author} {\bibfnamefont {D.~M.}\ \bibnamefont {Rayner}}, \bibinfo {author}
  {\bibfnamefont {D.~M.}\ \bibnamefont {Villeneuve}},\ and\ \bibinfo {author}
  {\bibfnamefont {P.~B.}\ \bibnamefont {Corkum}},\ }\bibfield  {title}
  {\bibinfo {title} {Alignment-dependent strong field ionization of
  molecules},\ }\href@noop {} {\bibfield  {journal} {\bibinfo  {journal} {Phys.
  Rev. Lett.}\ }\textbf {\bibinfo {volume} {90}},\ \bibinfo {pages} {233003}
  (\bibinfo {year} {2003})}\BibitemShut {NoStop}%
\bibitem [{\citenamefont {Shachar}\ \emph {et~al.}(2021)\citenamefont
  {Shachar}, \citenamefont {Kallos}, \citenamefont {De~Vries},\ and\
  \citenamefont {Bar}}]{Shachar_parker_valve_JPhysB_2021}%
  \BibitemOpen
  \bibfield  {author} {\bibinfo {author} {\bibfnamefont {A.}~\bibnamefont
  {Shachar}}, \bibinfo {author} {\bibfnamefont {I.}~\bibnamefont {Kallos}},
  \bibinfo {author} {\bibfnamefont {M.}~\bibnamefont {De~Vries}},\ and\
  \bibinfo {author} {\bibfnamefont {I.}~\bibnamefont {Bar}},\ }\bibfield
  {title} {\bibinfo {title} {A compact and cost-effective laser desorption
  source for molecular beam generation: comparison with simulations},\
  }\href@noop {} {\bibfield  {journal} {\bibinfo  {journal} {J. Phys. B}\
  }\textbf {\bibinfo {volume} {54}},\ \bibinfo {pages} {175401} (\bibinfo
  {year} {2021})}\BibitemShut {NoStop}%
\bibitem [{\citenamefont {Shlomo}\ and\ \citenamefont
  {Frumker}(2024)}]{Shlomo_In_situ_arXiv_2024}%
  \BibitemOpen
  \bibfield  {author} {\bibinfo {author} {\bibfnamefont {N.}~\bibnamefont
  {Shlomo}}\ and\ \bibinfo {author} {\bibfnamefont {E.}~\bibnamefont
  {Frumker}},\ }\href@noop {} {\bibinfo {title} {In situ measurement and
  control of the laser-induced strong field ionization phenomena}} (\bibinfo
  {year} {2024}),\ \Eprint {https://arxiv.org/abs/2401.17028} {arXiv:2401.17028
  [physics.atom-ph]} \BibitemShut {NoStop}%
\bibitem [{\citenamefont
  {S{\"a}ckinger}(2017)}]{Sackinger_transimpedance_amplifiers_book_2017}%
  \BibitemOpen
  \bibfield  {author} {\bibinfo {author} {\bibfnamefont {E.}~\bibnamefont
  {S{\"a}ckinger}},\ }\href@noop {} {\emph {\bibinfo {title} {Analysis and
  design of transimpedance amplifiers for optical receivers}}}\ (\bibinfo
  {publisher} {John Wiley \& Sons},\ \bibinfo {year} {2017})\BibitemShut
  {NoStop}%
\bibitem [{\citenamefont
  {Raizer}(1991)}]{Raizer_Gas_Discharge_Physics_Book_1991}%
  \BibitemOpen
  \bibfield  {author} {\bibinfo {author} {\bibfnamefont {Y.}~\bibnamefont
  {Raizer}},\ }\href@noop {} {\emph {\bibinfo {title} {Gas discharge
  physics}}},\ Vol.~\bibinfo {volume} {1}\ (\bibinfo  {publisher} {Springer},\
  \bibinfo {year} {1991})\BibitemShut {NoStop}%
\bibitem [{\citenamefont
  {Fitzpatrick}(2022)}]{Fitzpatrick_Plasma_Physics_2022}%
  \BibitemOpen
  \bibfield  {author} {\bibinfo {author} {\bibfnamefont {R.}~\bibnamefont
  {Fitzpatrick}},\ }\href@noop {} {\emph {\bibinfo {title} {Plasma physics: an
  introduction}}}\ (\bibinfo  {publisher} {CRC Press},\ \bibinfo {year}
  {2022})\BibitemShut {NoStop}%
\bibitem [{\citenamefont {Krausz}\ and\ \citenamefont
  {Ivanov}(2009)}]{krausz_atto_physics_Rev_Mod_Phys_2009}%
  \BibitemOpen
  \bibfield  {author} {\bibinfo {author} {\bibfnamefont {F.}~\bibnamefont
  {Krausz}}\ and\ \bibinfo {author} {\bibfnamefont {M.}~\bibnamefont
  {Ivanov}},\ }\bibfield  {title} {\bibinfo {title} {Attosecond physics},\
  }\href@noop {} {\bibfield  {journal} {\bibinfo  {journal} {Rev. Mod. Phys.}\
  }\textbf {\bibinfo {volume} {81}},\ \bibinfo {pages} {163} (\bibinfo {year}
  {2009})}\BibitemShut {NoStop}%
\end{thebibliography}%

\end{document}